# FPGA Based Real-Time Image Manipulation and Advanced Data Acquisition For 2D-XRAY Detectors

Wassim Mansour, Rattana Biv, Cyril Ponchut, Raphael Ponsard, Nicolas Janvier and Pablo Fajardo

*Abstract*—Scientific experiments rely on some type of measurements that provides the required data to extract aimed information or conclusions. Data production and analysis are therefore essential components at the heart of any scientific experimental application.

Traditionally, efforts on detector development for photon sources have focused on the properties and performance of the detection front-ends. In many cases, the data acquisition chain as well as data processing, are treated as a complementary component of the detector system and added at a late stage of the project. In most of the cases, data processing tasks are entrusted to CPUs; achieving thus the minimum bandwidth requirements and kept hardware relatively simple in term of functionalities. This also minimizes design effort, complexity and implementation cost.

This approach is changing in the last years as it does not fit new high-performance detectors; FPGA and GPUs are now used to perform complex image manipulation tasks such as image reconstruction, image rotation, accumulation, filtering, data analysis and many others. This frees up CPUs for simpler tasks.

The objective of this paper is to present both the implementation of real time FPGA-based image manipulation techniques, as well as, the performance of the ESRF data acquisition platform called RASHPA, into the back-end board of the SMARTPIX photon-counting detector developed at the ESRF.

## I. Introduction

IN the case of photons or other particle source facilities, a crucial step in the experiments is the proper detection of the produced particles after they go through some physical process. With the evolution of the technology of photon sources, the rate of produced and detected particles increases considerably [1]. It is important to mention that the ESRF machine has been upgraded to the 4th generation of synchrotron radiation, which means that the brilliance of the beam is multiplied by a factor of 100, and its coherency is multiplied by a factor of 30.

In other words, experiments that took days in the previous machine will now be achieved in minutes. More photons will make shorter exposure time and higher frame rates possible.

On the other hands, modern high performance 2D X-ray detectors are able to produce very high data rates in the range of 1 to 100 GB/s per megapixel. Such data streams are challenging to transfer, manipulate and process in acceptable times.

Real-time data manipulation, as early as possible to the detection stage, becomes now possible with the advance in technologies and the use of modern FPGAs. This will allow image reconstruction, rotation, accumulation and many other functionalities that are heavy tasks for CPUs to be performed at a very early stage even before transferring the data to the backend computer.

Similarly, the design and implementation of efficient data acquisition scheme with multi-gigabyte per second capabilities become a mandatory and unavoidable solution to deal with such data rates. RASHPA (RDMA-based Acquisition System for High Performance Applications) is a generic data acquisition framework developed at the ESRF and presented in [2] and [3]. It is optimized for the transfer of 2D detector data, i.e images, metadata, etc. It relies completely on Remote Direct Memory Access (RDMA) mechanisms. RASHPA is able to push data, at very high speed into the address space of one or several backend computers. The scheme provides a high standardization level in the data transmission pipeline from the detector up to the software application for further processing, visualization or storage.

This paper details the whole data acquisition chain for the SMARTPIX 2D-XRAY photon-counting detector developed at the ESRF [4]. The paper is organized as follows: section II briefly presents the SMARTPIX detector and provides a description about the front-end implementation as well as the backend requirements. Section III shows the hardware implementation of many image manipulation algorithms such as image rotation, accumulation, spectroscopic mode and image reconstruction for different configurations. Section IV, presents the data acquisition platform, RASHPA. Section V discusses the obtained results. Finally, section VI concludes and provides some perspectives.

W. Mansour is with the European Synchrotron Radiation Facility (ESRF), Grenoble, France. (e-mail: mansour@esrf.fr)

R. Biv is with the European Synchrotron Radiation Facility (ESRF), Grenoble, France. (e-mail: biv@esrf.fr)

C. Ponchut is with the European Synchrotron Radiation Facility (ESRF), Grenoble, France. (e-mail: ponchut@esrf.fr)

R. Ponsard is with the European Synchrotron Radiation Facility (ESRF), Grenoble, France. (e-mail: ponsard@esrf.fr)

N. Janvier is with the European Synchrotron Radiation Facility (ESRF), Grenoble, France. (e-mail: janvier@esrf.fr)

P. Fajardo is with the European Synchrotron Radiation Facility (ESRF), Grenoble, France. (e-mail: fajardo@esrf.fr)

## II. SMARTPIX PHOTON DETECTOR

SMARTPIX[4] is a scalable and versatile pixel detector system with minimized dead areas and with energy resolving capabilities based on Medipix3RX readout chip [5] and active edge pixel sensors.

Medipix3RX readout ASIC is a 256x256 photon-counting pixel matrix with 55x55 µm$^2$ pixels. In addition to conventional photon-counting, Medipix3RX is capable of reconstructing the charge signal of one X-RAY event from charge splits in neighboring pixels, resulting in more accurate threshold discrimination and better spectroscopic capabilities without loss of spatial resolution. Medipix3RX also provides an arbitrated mode which increments only the counter of the pixel receiving the highest charge share of an X-RAY hit, thereby eliminating count multiplicity due to charge sharing.

Medipix3RX contains two 12-bits counter configurable to support: single counter or continuous readout modes. In the single counter mode, only one counter is exposed to photons before it is read, while in continuous readout mode, one counter is exposed while the second is being read. Those counters have a configurable pixel size of (1-bit, 6-bits and 12 bits mode). In addition to that, the Medipix3RX chip support pixel size of 24-bits by concatenating both 12-bits counters. This provide the possibility to have a bigger exposure time but limits the readout mode as both counters are exposed to photons at the same time and readout together.

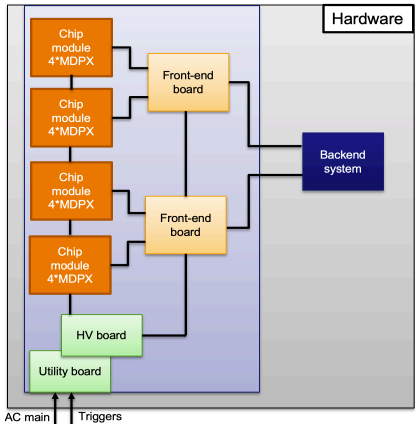

Fig. 1. Block diagram of the SMARTPIX detector

A complete SMARTPIX detector consists of four modules (four Medipix3RX chips each) connected to two Kintex7 FPGA-based front-end electronic boards responsible for the data readout. These boards are powered and triggered by the so-called utility board. Each of the front-end boards is connected, via a fiber data link running at 40 Gbps, to an FPGA based back-end board responsible for the data acquisition process. Since SMARTPIX is capable of producing 6000 frames per second at a rate of 22 Gbps, the 40 Gbps link is able to absorb all the produced frames without any bottleneck.

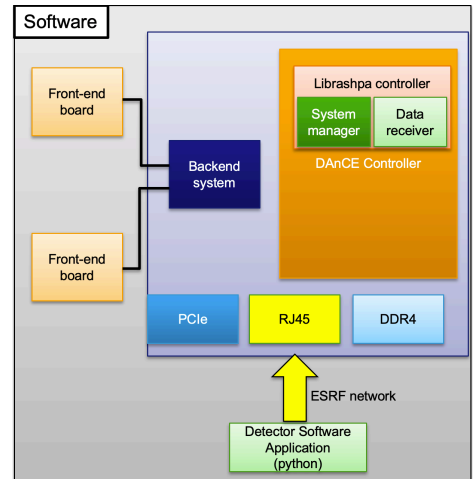

Fig. 2. Software architecture of the SMARTPIX detector

Each back-end board is connected to the backend computer to which the final data will be injected through a PCIe Gen3x16 (Peripheral Component Interconnect express) connecters [6].

A SMARTPIX controller that acts as a driver to the backend, front-end and MEDIPIX chips is developed installed on the backend computer. This controller embeds another controller for RASHPA data acquisition platform called LIBRASHPA controller. SMARTPIX controller allows any PC connected to the ESRF network to configure have access to the detector hardware. Fig. 1 shows the block diagram of the hardware architecture of a 1 MegaPixels SMARTPIX system however the software architecture is shown in Fig.2.

The front-end boards, embed a Xilinx kintex7, XC7K325T, FPGA. These boards are responsible for controlling and configuring the Medipix3RX chips. In addition to that, they perform the data readout from one, four or eight Medipix3RX chips. The readout is done via up to eight read channels per chip at 300 MHz clock frequency. Parallel bitwise readout data from the Medipix3RX chips are used to reconstruct the pixels, then interleave them together and transmit them to the backend board through the optical link via a Xilinx-based AURORA protocol.

The backend board, an ALVEO U200 [7], based on Xilinx Ultrascale+ technology, is responsible with the help of a SMARTPIX controller, to transmit the control commands to the front-end boards and receive the data streams when configured in acquisition mode. The backend boards also perform the following functionalities: de-interleaving the data streams with respect to the number of configured readout chips (1, 4 and 8 chips are supported); performing real-time static image rotation (0, 90, 180, 270 degrees) per chip, module or global frame and per pixel size; accumulate n-frames; combine counters to generate 24 bits pixel mode; spectroscopic mode (splitting one image into four sub-images when dealing with a non-monochromatic beam with different threshold per pixels), image and frame reconstruction and dummy-pixel suppression (pixels between chips, modules and systems).

In addition to these functionalities, the backend board integrates RASHPA, the ESRF data acquisition platform based on RDMA that allows data to be injected directly following pre-configured rules into the DDR of the data receiver without any CPU intervention. Together, the hardware RASHPA IP and the dedicated software (LIBRASHPA) are put in place to perform the data acquisition process for the SMARTPIX detector.

## III. Hardware Implementation

The focus of this paper will be on the hardware implementation of the different real-time image manipulation functionalities on the back-end FPGA.

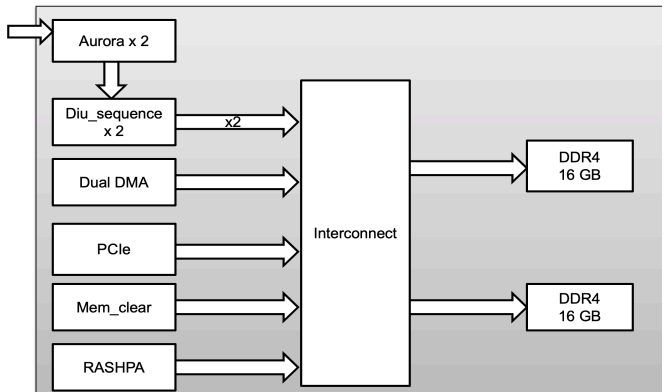

Fig. 3. Block diagram of the backend architecture

As shown in Fig. 3, the backend FPGA implements two Aurora for interfacing with the front-end boards. These interfaces are responsible for the SMARTPIX configuration and data acquisition.

Received data will then enters two parallel pipelined blocks called *diu-sequence* representing each front-end board. These *diu_sequence* blocks are responsible for all the real-time image manipulation functionalities and the data storage into the on-board DDR4. In addition to that, there is an IP called *memory initializer* that sets to a configurable value, all the addresses in the DDR4 when the FPGA is powered up. This is a crucial step to start a clean acquisition and manipulate the data correctly in the DDR.

Finally, a RASHPA block is implemented to perform the data acquisition. It will select the data to be sent to the back-end computer based on some rules set by the user. In fact, it will activate two Central Direct Memory Access IPs (CDMA), in a ping-pong technique and configure them to send the requested data from the DDR into the destination via a PCIe interface.

It is important to note that these IPs are interconnected between each others via an AMBA AXI protocol (Advanced Microcontroller Bus Architecture/Advanced eXtensible Interface) [8][9].

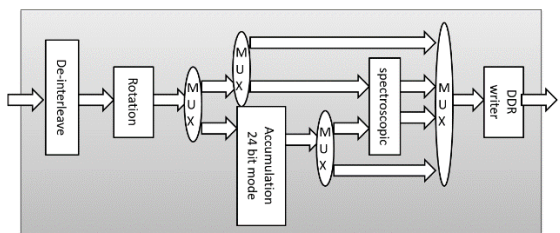

Fig. 4. Architecture of the detector interface sequence

The *diu_sequence* block is the core block responsible of the manipulation of the images at real-time. This block is composed of a data de-interleaver, a rotator, an image accumulator/24-bits concatenator mode, a spectroscopic splitter and a ddr_writer block as shown in Fig. 4.

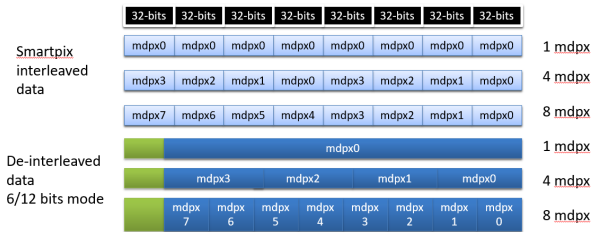

Fig. 5. Interleaved and de-interleaved data based on the number of readout chips and the pixel size.

The de-interleaver block is responsible to split out the interleaved data streams coming from the front-end board into up to eight streams representing each, data corresponding to one Medipix3RX chip. This de-interleaver has to take into account the user configuration including the number of readout chips under acquisition, and the size of the pixels. Every three streams of 256-bits interleaved pixels are transformed into four streams of 256-bits which represent an integer number of pixels after transforming the 6-bits mode into 8-bits and the 12-bits mode into 16-bits in order to satisfy an important RASHPA requirement which is byte alignment.

SMARTPIX has also a special greyscale one bit mode, for which data are kept in 256 bits, where every bit represents a pixel. In order to simplify the hardware implementation for this specific mode, a different de-interleaver data path is implemented.

Fig. 5 illustrates the output of the de-interleaver block based on the pixel size and the number of readout chips under acquisition.

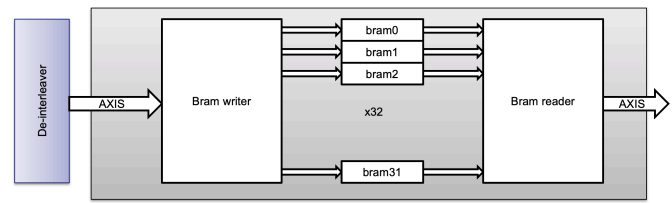

Fig. 6. Block diagram of the rotation IP

Next block is a rotation IP. The rotation algorithm is invented in the frame of this project. It is based on three main sub-blocks: a memory writer, 32 Block-rams and a memory reader.

The writer is responsible to write the interleaved data into 32 parallel block-rams of total capacity equal to two frames of eight Medipix3RX chips when operating at 16-bits per pixel which is 128 kbits each. This is chosen because the writer should write an image while a reader is reading the previous one in a ping-pong technique. The write process should place each pixel into its correct location with respect to the rotated readout frame. To do so, the writer is configured with a starting address that depends on the rotation angle. With every SMARTPIX data transfer this starting address has to be shifted either backward or forward by the number of Medipix3RX chips under acquisition. This allows to separate the de-interleaved data in the parallel block rams.

The writer block takes care of the number of configured readout chips, the size of the pixel and the angle of rotation per chip. It supports four rotation angles (0, 90, 180 and 270 degrees) for the 6 and 12 bits mode, and two rotation angles (0 and 180 degrees) for the 1-bit mode. The process is fully pipelined with a latency of one image.

The reader block then flushes the fully rotated image from the block-rams, at a rate of 256-bits per cycle. The output of the reader is a 256-bits wide AXI-stream interface.

After rotating the frame, the data have the opportunity either to perform an accumulation of n frames (where n is a configured value) or 24 bits mode, or to be splitted out into four sub-images (spectroscopic mode, 110x110 µm2 pixels) or do either accumulation/24-bits and spectroscopic mode together. Implementation of these functionalities require the use of several FPGA memories. The ultrascale+ provides Ultra-Rams (URAM) blocks that fit very well for this kind of implementation.

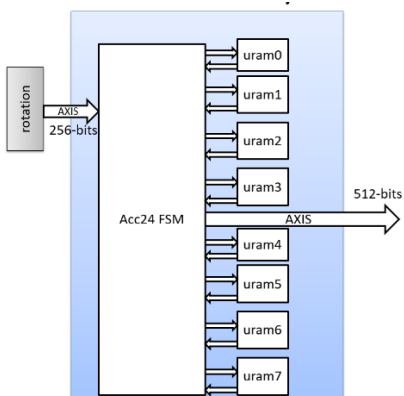

Fig. 7. Block diagram of the accumulation/24-bits mode IP

For the accumulation and 24-bits mode, the same hardware is used altogether with eight URAMs. It is important to note that 16-bits pixels are received from every frame, resulting 32-bits pixel after accumulation or shifting. A block called Acc24 is implemented. It is responsible to write one frame in the URAMs, and while receiving the second frame, it reads the previously stored frame, then it performs the arithmetic operation. If the chosen arithmetic is a shift operation (24-bits mode), the resulting data are flushed out via the 512-bits AXI stream interface. However, if the arithmetic is an addition (accumulation mode), the pixels are summed up, and stored back in a different location within the URAMs. This allows parallelizing all the operations, thus maximizing the throughput. Fig. 7, shows a block diagram of the accumulation/24 bits hardware implementation.

The spectroscopic is an operation mode in which each adjacent pixel corresponds to a different frame or sub-images as shown in Fig. 8.

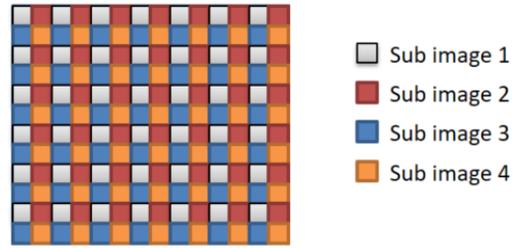

Fig. 8. Frame splitting in spectroscopic mode

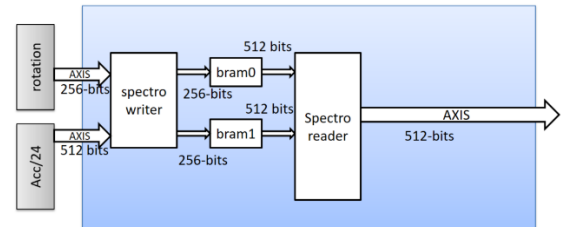

Fig. 9. Block diagram of the spectroscopic mode IP

In order to implement the spectroscopic mode, the same concept as for the rotation IP is applied. A writer, reader and two dual-ports BRAMS have to be used. The writer split into even and odd pixels of a line into two BRAMs. Whenever a line is split out, the reader will flush it out, while the writer is splitting the next line adding an extra latency of one line without affecting the throughput. The hardware architecture of the spectroscopic mode is shown in Fig. 9.

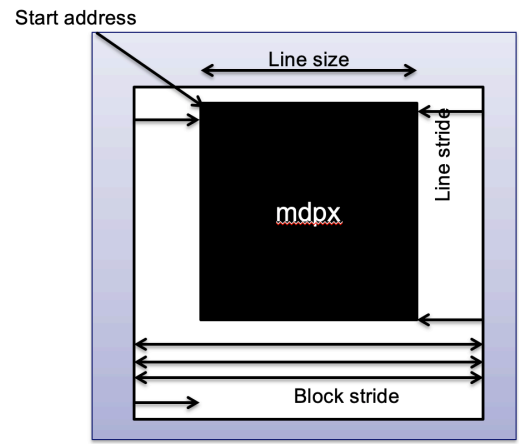

Fig. 10. DDR Writer basic parameters

The last IP in this chain is called DDR_writer. It is responsible for placing each line within the frame into its correct address in the DDR. Given a start address in the DDR, a line size in bytes, a line stride (how many bytes one needs to skip between two lines) and block stride (how many bytes one needs to skip between two Medipix chips), Fig.10., allows a full reconstruction of the SMARTPIX frame per module or per system. It also allows the insertion of dummy pixels, which are null pixels between the readout chips or null pixel between SMARTPIX modules.

## IV. RASHPA

RASHPA allows detectors to push data (images, regions of interest (ROI), metadata, events etc...) produced by 2D X-ray detectors directly into one or more backend computers. RASHPA's main properties are its scalability, flexibility and high performance. It is based on RDMA (Remote Direct Memory Access) mechanism to provide the maximum possible bandwidth per link. It is also intended to have an adjustable bandwidth that can be compatible with any backend computer.

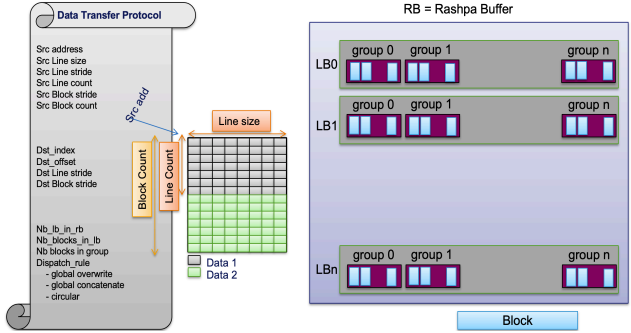

Fig. 11. RASHPA data transfer rules

In order to send perform the RDMA transfer from the backend FPGA to the backend computer, the destination should allocate some buffers to which the date will be sent as shown in Fig. 11. These buffers are called Local Buffers (LB), to which a group of images can be sent. All the allocated local buffers are grouped in the so-called RASHPA buffer (RB).

In the data acquisition process, the user specifies a set of rules, that RASHPA will follow in order to operate. These rules concern both the source (FPGA DDR) and the destination (backend computer buffers). Concerning the source, RASHPA should know the source address, the size of the line to be transferred, the line stride, line count, block stride and block count.

Regarding the destination rules, RASHPA should know the index of the local buffer in RASHPA buffer, the offset in bytes with respect to the local buffer base address, the destination line stride and the destination block stride.

In addition to that RASHPA should also be configured with a set of parameters to allow different dispatching rules (global overwrite, global concatenate or circular modes). These parameters are the number of local buffers in a RASHPA buffer, number of blocks in a local buffer, and number of blocks in a group.

RASHPA hardware is based on the concept of data channels. These data channels acting as Central Direct Memory Access (CDMA) configurators. The user defined rules will be used by these channels to select the data stored in the DDR by the DDR writer. These rules will then be transformed to DMA configurations, which activate an AXI Central Direct Memory Access (CDMA IP), implemented in the FPGA. The AXI CDMA reads the data from the DDR and sends them to the data link.

RASHPA is able to coop with different kind of data links, in this work the used data link is selected to be a PCIe.

## V. EXPERIMENTAL RESULTS

In this section are shown the experimental results obtained when configuring the SMARTPIX detector to send the ESRF logo, and the back-end board to support several acquisition parameters. The objective here is not to show the throughput rather than the functionalities, knowing already that the throughput limit is the readout bandwidth of the Medipix3RX chips.

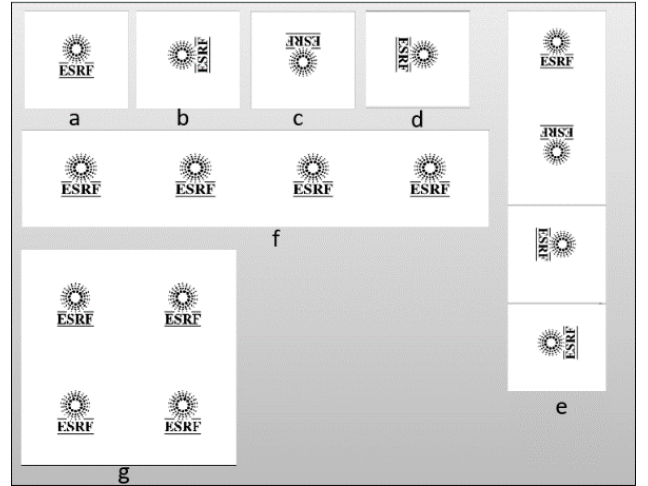

Fig. 12. Obtained results when using several configuration modes

In the absence of XRAY sources, the software controller programs the Medipix3RX chips with a pre-defined frame. As mentioned earlier, the selected frame is the ESRF logo. Fig. 12 shows outputs of several operation mode. (a) shows the acquired image where 1 Medipix3RX chips is selected to operate at 12-bits mode pixel size with no rotation. (b) rotation 90 degrees, (c) rotation 180 degrees and (d) rotation 270 degrees. (e) shows the image obtained of 4 chips each one with a different rotation angle, (f) shows a reconstructed image of four Medipix3RX in a row. And (g) shows result of the spectroscopic mode reconstructed in the same frame.

TABLE I
RESOURCE UTILIZATION OF THE IMPLEMENTED HARDWARE

| Resource | Utilization | Available | Available (%) |
|---|---|---|---|
| LUT | 320214 | 1182240 | 27.09 |
| LUTRAM | 16441 | 591840 | 2.78 |
| FF | 373458 | 2364480 | 15.79 |
| BRAM | 1662.50 | 2160 | 76.97 |
| URAM | 256 | 960 | 26.67 |

Table I, shows the resource utilization of the hardware implementation on the ALVEO, U200 board. The main conclusion that can be extracted from the table is the massive use of block-rams, and this is due to the rotation algorithm that is generic and parallelized. The high usage of block rams means that they will be distributed in the FPGA. This distribution leads to a routing problem caused by the net delays and clock skewing, so a lot of effort was injected in this topic to overcome the situation by placing and routing correctly the block rams of one *diu_sequence* IP within one silicon region (SLR) of the FPGA.

## VI. CONCLUSION AND PERSPECTIVES

In this work, the implementation and results of real-time image manipulation and RASHPA PCIe based data acquisition for the SMARTPIX detector are detailed.

The results obtained when configuring the detector to several mode of operations shows the capabilities of the algorithms and implementation as well as the performance of modern FPGA technologies.

Future work will include adding more functionalities to the system and using partial reconfiguration to minimize build time and FPGA resources.

On the other hand, RASHPA over RoCE-V2(RDMA over Converged Ethernet) will be a main focus in future work.